# Information storing by biomagnetites

[1]Istvan Bokkon, [2,3,4]Vahid Salari

[1]*Doctoral School of Pharmaceutical and Pharmacological Sciences, Semmelweis University, Hungary*
[2]*Department of Physics, Shahid Bahonar University of Kerman, P. O. Box 76175, Kerman, Iran*
[3]*Afzal Research Institute, Kerman, Iran*
[4]*Kerman Neuroscience Research Center, Kerman, Iran*

**Abstract** Since the discovery of the presence of biogenic magnetites in living organisms, there have been speculations on the role that these biomagnetites play in cellular processes. It seems that the formation of biomagnetite crystals is a universal phenomenon and not an exception in living cells. Many experimental facts show that features of organic and inorganic processes could be indistinguishable at nanoscale levels. Living cells are quantum "devices" rather than simple electronic devices utilizing only the charge of conduction electrons. In our opinion, due to their unusual biophysical properties, special biomagnetites must have a biological function in living cells in general and in the brain in particular. In this paper we advance a hypothesis that while biomagnetites are developed jointly with organic molecules and cellular electromagnetic fields in cells, they can record information about the Earth′s magnetic vector potential of the entire flight in migratory birds.

**Key words**: biomagnetite formation by biological control, Aharonov-Bohm effect

# 1. Introduction

Usually when we hear about crystals in relation to the living organisms or cells, we first think of pathological cases such as kidney stones, gallstones, etc. Deficient biochemical processes or microorganisms can produce different kinds of pathological biocrystal formations [2, 11]. However, some kinds of biocrystals can have functional roles in living cells.

Many scientists treat biomagnetites in a very cursory fashion while others use biological self-assembling techniques to create inorganic biomagnetites within living bacterial cells, because these biomagnetites have such special features that make them capable of functioning in computers as information storage components [40]. If biomagnetites can work in computers, why can't they work in living cells as well?

It has been well accepted by now that different cells of many species contain inorganic biomagnetic crystals. Joseph Kirschvink, at California Institute of Technology, researched for many years various types of living creatures such as bacteria, insects, birds, fishes, etc. and has found that biomagnetites play an important role in their spatial orientation [32, 36]. He has also found that biomagnetic $Fe_3O_4$ iron-oxide crystals in the human brain are composed of 50-100 granules. According to his calculations there are five million magnetic crystals in every gram of human brain, their size being on the order of 10-100 nm [31, 33]. Considering the mass of the whole human brain, we may assume the presence of a billion nanocrystals in our brain. According to Kirschvink's experiments, there are functional connections between biomagnetites and organic molecules. Besides, the assumption that biomagnetites not only perceive magnetic fields, but also can take part in the inheritance of magnetosome polarity was raised a long time ago [28,48]. Here, we raise the possibility of having heredity in which information is not contained fully at the DNA level. In this connection, we note that the process of information storage occurs in biomagnetites and not in DNA, but this fixed information could be manifested at the DNA level.

# 2. Magnetobiology and Associated Problems

When we talk about magnetic effects on humans, two different magnetic field ''types'' are commonly distinguished: (1) a static magnetic field, which exists in the region of a large magnet; and (2) a magnetic field which is pulsed at frequencies higher than 10 Hz, often abbreviated as EMF (electromagnetic fields). The area of investigation of these special effects is termed ''magnetobiology'', some sub-fields of which are extremely contentious, while others have already been established in medical applications. Nonetheless, there are some basic and all together paradoxical problems in magnetobiology. For example, how can weak magnetic fields (e.g., Earth's magnetic field), or, ultra-weak electromagnetic expositions have effects on cellular processes? On the other hand, numerous scientists agree that static magnetic fields of up to 10 Tesla have no obvious effects on long-term plant growth, mouse development, body temperature, or brain activity [5, 6, 45] to name a few biological phenomena of interest. Also research indicates that humans are sensitive to small changes in magnetic field gradients, but not to the overall magnetic field [54].

There exists some strong interaction which results in the parallel alignment of all atomic magnetic moments inside the body. The magnitude of this spontaneous magnetic moment decreases if the sample temperature increases, because thermal fluctuations act against any order due to magnetic interactions [3]. Why do these random disturbances not destroy the magnetobiological effects? At the same time, how can living cells resist or compensate the strong effects of natural and artificial magnetic or electromagnetic environments?

According to [1], the mean noise potential energy without the presence of magnetic fields is $U_{noise} = \frac{1}{2}\kappa\theta_{kT}^2 = kT$, where $\kappa$ is a scalar coefficient, $\theta_{kT}$ is the angular displacement of a biomagnetite element which is placed in the cell and is free to rotate in that place and generate biological effects through that rotation, $k$ is Boltzmann's constant and $T$ is the absolute temperature. If this rotation affects biology, the magnetite system must be coupled to an element through some harness so that motion of the magnetite system changes the biologically sensitive conformation of that element. That transition must require an energy in excess of $kT$, if it were not to take place regularly through thermal agitation without any coupling to magnetic fields. Now we want to compare the energy of a magnetic field to thermal noise energy $kT$. Suppose that the magnetic field is in the form of $B(t) = B_0 \cos\omega t$. We denote here $\mu$ for the magnetic moment of biomagnetite element. The total inserted torque on the biomagnetite element is obtained from a linear equation as below [1],

$$T = I\ddot{\theta} = -\beta\dot{\theta} - \kappa\theta + B_0\mu\cos\omega t\cos\phi\sin\xi + \chi(t), \tag{1}$$

where $\beta$ is a scalar coefficient, $\theta, \dot{\theta}$ and $\ddot{\theta}$ are angular displacement, angular velocity and angular acceleration, respectively. $\phi$ is the angle between the direction of the applied torque and the final acceleration, and $\xi$ is the angle between the applied field $B$ and the magnetic moment of biomagnetite $\mu$, and $\chi(t)$ represents the effects of thermal agitation. Because of the linearity of equation (1) the magnetic and thermal noise amplitudes at every time instant can be written in the form of $\theta(t) = \theta_B(t) + \theta_{kT}(t)$, where $\theta_B(t)$ is the solution without the presence of noise, thus $\chi(t) = 0$, and $\theta_{kT}(t)$ is the solution without a magnetic field which means $B(t) = 0$. The general form of the solution for $\theta_B(t)$ is $\theta_B(t) = \theta_0 \cos(\omega t + \phi)$ where $\theta_0$ is written in the form of equation (2),

$$\theta_0^2 = \frac{(B_0\mu)^2}{I^2(\omega_0^2 - \omega^2)^2 + \beta^2\omega^2} \tag{2}$$

One can define two important energies here: $E_T = \frac{1}{2}I\omega^2\theta_{rms}^2$ which describes the interaction energy between the magnetic field and the biomagnetite element, and $E_V = \frac{1}{2}\kappa\theta_{rms}^2$ which is the representative of potential energy of the biomagnetite. In the above energy relations, $I$ is the moment of inertia of the magnetic element due to $B(t)$, and $\theta_{rms}^2$ is obtained from $\theta_{rms}^2 = \frac{\theta_0^2}{2}$. If we let $\omega_0^2 = \frac{\kappa}{I}$, then the ratio $\frac{E_V}{E_T} = \frac{\omega_0^2}{\omega^2}$ can be interpreted for different values of variables. The energy of interaction, $w$, between the magnetic field, $B$, and the magnetic moment, $\mu$, is $w = \vec{B}\cdot\vec{\mu}$. For the Earth's field, $\omega = 0$ and $B_{earth} \approx 50\mu T$, we have $E_v = B_{earth}\mu \succ\succ kT$ for biomagnetites with large net magnetic moments. We also have similar result for MRI magnetic field intensity in the 1-14 Tesla range, although it is much stronger. Considering small values of $\omega$, and neglecting $\omega^2$ in the term $\omega_0^2 - \omega^2$ in equation (2) and bearing in mind the ratio $\frac{E_V}{E_T}$, we have $E_v \prec \frac{(B_0\mu)^2}{4}\frac{1}{\kappa(1+\beta^2)}$ . Now, if we let $\frac{\beta}{\kappa} = \frac{1}{\omega}$ we have $E_v \prec \frac{B_0^2 \mu^2}{8\beta\omega}$. For any magnetite and requiring that the magnetic fields must not affect biology $E_v \prec\prec kT$. By way of the required information in the above relationships there is the possibility to investigate the effects of different magnetic fields on biomagnetites. The evoked human brain activity has $\leq 10^{-13} Tesla$ intensity, and the magnetic fields due to spontaneous currents in the brain are about $10^{-12} Tesla$, while the earth magnetic field is $\geq 10^{-5} Tesla$ [3].

There are many models of weak magnetic effects on living cells. For example, with the help of biomagnetites, Eddy electric currents, classical and quantum oscillator models, cyclotron resonance, interference of quantum states of bound ions and electrons, coherent quantum excitations, parametric resonance, stochastic resonance, bifurcation, magnetosensitive free-radical, etc. [9]. Adair [1] has shown that the energies transmitted to the magnetite elements by fields less than $5 \times 10^{-6} Tesla$ will be much less than thermal noise energies. Thus, the effects of such weak fields will be overwhelmed by thermal noise and cannot be expected to affect biology.

The basic question is "*why these models could not solve the basic problem of magnetobiology?*" Perhaps, there are not real contradictions among the above mentioned models, because, living cells use several kinds of information processes simultaneously, but these processes are too complicated to find relations among them.

## 3. Electric, magnetic, electromagnetic and acoustic signals produced by living cells

- Living cells can generate and use electric, magnetic and electromagnetic (both coherent and incoherent biophotons) waves and also acoustic waves as conformational changes in macromolecules called conformons (analogous to lattice vibrations of phonons in solid crystals) [10, 15, 23, 37, 49, 51, 52]. Electricity is a basic trait of cells, because all "biomolecules" are ions or biomolecules which are endowed with high electric dipole moments. When their charges move, an electromagnetic field is generated. Magnetic features can emerge from free radicals, organic molecules with metals or biomagnetites [26]. In addition, according to ESR (electron-spin resonance) experiments, living cells or organisms can have paramagnetic features in their native states [59]. Here, we note that conformons originate from continuously moving molecules.

- Different parts of cells (such as DNA, RNA, proteins) show piezoelectric and semiconductor properties [19, 24, 56, 60]. The piezoelectric effect refers to that property of matter, which can convert electromagnetic oscillations to mechanical vibrations and vice versa, or electric oscillations to mechanical vibrations and vice versa. The piezoelectric property of cells can produce circular polarized light pulses which indicate that living molecules are not raceme mixtures of optically active molecules. Organic semiconductors have crystal-like structures and electrical conductivity as diodes. The electric fields of a wave can couple to the mobile carriers within a semiconductor structure and modify its electronic and elastic properties. Optical signals (biophotons) can be stored by the surface acoustic waves (conformons as mechanical vibrations in macromolecules) in the semiconductor in a photon-atom-bound way [31], and can be re-assembled into light after very long delay times and at a remote location of the sample.
- The membrane lipid has a density of about $10^{11}$ pores m$^{-2}$, and it is not in a uniform non-conductive bulk phase [13]. It is a quasi insulator, with conductive and non-conductive parts, and also there are semiconductor proteins in it. Finally, we note that the current-voltage relationship has a nonlinear characteristic.
- Cytoplasm is not simply an aqueous solution of macromolecules, but is a structurally and dynamically organized network (cytoskeleton) of interconnected (semiconductor) protein polymers in the ordered water/ion solution. Namely, living cells exhibit a liquid-crystal like state. [29, 12]. Liquid crystals show some of the orientational order of a solid, but the molecules are mobile. The regulated living structures can produce coherent or laser-like oscillations, which are called Fröhlich oscillations [22].
- It seems that the coherence – in the ultrashort time – is a universal phenomenon and not limited to biochemical processes [62]. Many processes work on the ultra short femtosecond time scale, which appears to avoid the effects of thermal noise or fluctuations and also can produce coherent biophotons.
- Living systems show fractal features. Fractal systems, which are the best models of living and non-living world, have special features. Namely, fractals can resist strong forces at the same time, they can use very weak ones [16, 42, 64]. Phase delay is a very important property of fractals, which can be generated by the noise of non-linear systems. Namely, properties of fractals can be related to the problem of useful noise [46].

- Living cells and organisms can use nonlinear creative white noise for signal amplification. In accordance with experiments, random noise can help neurons to react (as a non-linear resonance) against weak signals [47].

In brief: living cells can generate and use electric, magnetic, electromagnetic and acoustic waves (conformons), and convert them from one form to another.

**4. Ion gate model of biomagnetites**

What are biomagnetites doing in various cells? In the case of bacteria the answer may be simple. Magnetite crystals can take part in navigation processes and sense Earth's or other magnetic fields, an effect called magnetotaxis. But in the case of birds the answer is not simple at all. How can a migratory bird find its way home from 3000 kilometers away with the help of biomagnetites? Moreover, what are billions of magnetites doing in the human brain?

Kirschvink has proposed a model that biomagnetites can open or close ion gates in the cells while they perceive the Earth′s magnetic fields or different electromagnetic fields [34]. This idea was encountered with many contradictory opinions, doubting that the Earth′s magnetic field would be strong enough to generate significant biological effects on biomagnetites.

According to Zeilinger, "Yet I am not convinced that living systems are just classical machines" [39]. However, living cells are quantum "devices" rather than simple mechanical and electrical machines. Many experiments show that ultra weak forces work - fast and accurately - in cells, as well as information or regulation processes are much faster than molecular processes. If ultra weak processes do not work accurately and very fast (at femtosecond time scales) in cells, different effects of natural and artificial environment can disintegrate these processes. Of course various forms of natural and artificial radiation can have effects on the cellular processes but cells can compensate for this within a wide range. As we mentioned before, living cells can sense and use the Earth's weak magnetic fields, which are 100 billion times stronger than brain's magnetic processes [27]. At the same time, during MRI experiments we are exposed to very strong 1-14 Tesla magnetic fields, which are much stronger than Earth's magnetic field. If biomagnetites work as a regulator of ion gates, during MRI scans, magnetic radiation can have strong effects on biomagnetites which can be arranged in one direction and thus the brain's processes can be collapsed. As a result, an ion gate model is hardly possible.

However, there exist additional biophysical possibilities as biomagnetites can take part in information processes in cells. We believe that it is insufficient for a migratory bird to find its way home by simply perceiving the Earth′s magnetic field. The migratory bird has to record the magnetic map or magnetic vector potential map of the entire journey. But how can a migratory bird find the exact way home, to its nest under the certain eaves of a certain house in a certain city from the distance of some thousand kilometers. Here we suggest that while biomagnetites are built up jointly with organic molecules and cellular electromagnetic fields in cells, these biomagnetites can record information of the Earth′s magnetic vector potential during the entire flight.

**5. Biomagnetite crystal formation by biological control**

In previous sections we could see that living cells can create and use weak electric, magnetic, electromagnetic and acoustic waves, and convert one of them into another. Electrons play a very important role in the information flow between organic and inorganic materials in various cells. Ge et al. - at Lawrence Berkeley National Laboratory - examined the behavior of electrons at interfaces [25]. A piece of inorganic silver was coated with organic paraffin and it was illuminated with a tunable laser by a femtosecond pulse. The electrons came out from the silver surface and could bind to the lattice of organic paraffin as polarons. The polaron existed for 1000 femtoseconds and came back via a tunnel-effect to silver. As the authors emphasized, this phenomenon is very important in biochemical processes, namely a functional electric connection can exist between organic and inorganic materials. Since biomagnetites are in connection with surrounding organic protein molecules, these molecules can regulate the development of biomagnetites by electric operating processes.

However, bioelectromagnetic forces can regulate the formation of biomagnetites in cells as well. Electromagnetic waves mostly affect the length of cell membranes which are in a functional connection with biomagnetites [50]. In *in vitro* experiments weak electromagnetic fields exert a direct influence on the kinetics of crystal formation [8]. Since living cells can produce coherent electromagnetic waves (biophotons) [4, 7, 38, 53], these cellular electromagnetic forces can also regulate the formation of biogenic magnetites. Namely, holographic lithography-like mechanism could work within cells (formation of biocrystals by interference of non-coplanar coherent biophotons) [14].

Ursula Liebl et al. proved evidence – with femtosecond spectroscopy – for driving of a reaction in a protein complex by coherent motions, and suggested the functional importance of coherent vibrations operating on a femtosecond timescale [41]. It seems that the ultrashort time coherence is a universal phenomenon in biochemical processes, thus femtosecond processes can produce coherent vibrations operating at an interface between organic molecules and inorganic crystals. At nano-levels there are unusual magnetic and electric nonlinear fluctuations, which make modeling relatively difficult. Because of quantum size effects, matter at the nanometer scale has very special properties; altered thermodynamics and modified chemical reactivity.

However, biomagnetic crystals are solitary and structurally well ordered magnetic domains with stable magnetized and the maximum magnetic moment per unit volume required for magnetite [35, 43]. Biomagnetite crystal morphology is cubo-octahedral with the {111} direction (see Figure 1) which yields unusual particle shapes in geological magnetite crystals, so the production of this biomineral must be under precise biological control. This biological control of biomagnetite formation can be achieved by surrounding organic protein molecules and coherent electromagnetic fields (biophoton) in cells.

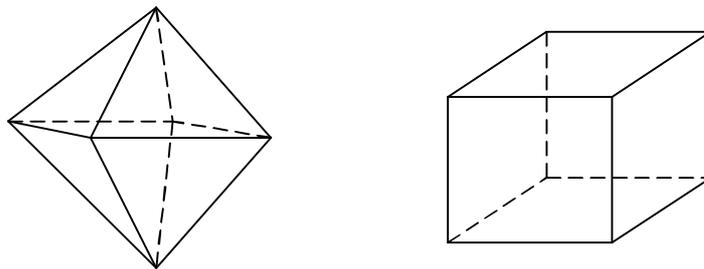

**Fig. 1**. Crystal morphology from left to right: octahedral {111} and cubic {100} forms.

We should see whether the nanometer scale is the scale at which soft and hard materials sciences overlap.

## 6. Information storage in biomagnetites and the Aharonov-Bohm effect

Quantum physical Aharonov-Bohm effect was proven via experiments on electron interference [30]. The essence of this effect is seen as follows. If we take a piece of static magnet, its static magnetic field is shielded but an effect still exists which can change the wave phase of electrons (see Figure 2). This effect is more fundamental than a magnetic feature called the magnetic vector potential.

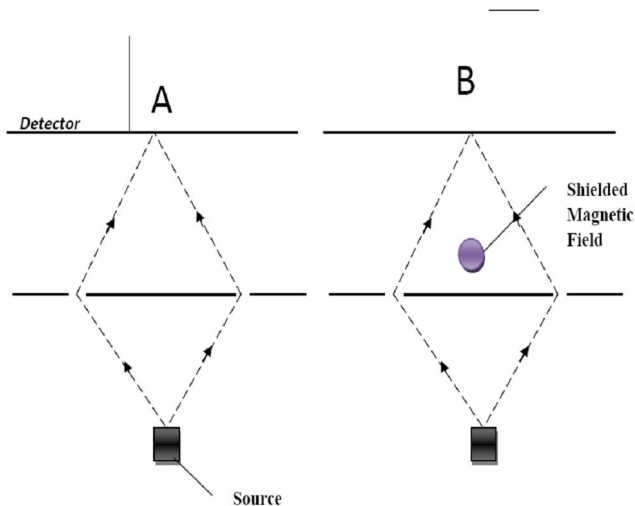

**Fig. 2**. Aharonov-Bohm effect: In the interference experiment for electron particles, in part A there is no any magnetic field around the set-up, but in part 2 there is a shielded magnetic field after the two-slit which has no any classical effect on the paths of electrons. The interference representations on detectors A and B are different. The magnetic field in part 2 has changed the wave phase of electrons.

Here we suggest that spin-modulated resistance could work in biomagnetites through the Aharonov-Bohm effect. In this effect, the phase of the electron wave depends on the magnetic vector potential, which causes a phase difference and interference between partial waves. Through the Aharonov-Bohm effect weak geomagnetic fields can have effects on living cellular processes.

Biomagnetites can also be viewed as so-called ferromagnets. Manyala et al. argued that magnetoresistance which can rise from different mechanisms in certain ferromagnets is a quantum interference effect [44]. In addition, Tsukagoshi et al., in their experiments, reported that spin-polarized electrons can be injected into non-ferromagnetic materials (multi-walled carbon nanotubes) from a ferromagnet, finding direct evidence for the coherent transport of electron spins [61]. The above-mentioned mechanisms can also work in cells. Namely, biomagnetites can take part in information storage and operating processes in cells through the Aharonov-Bohm effect.

Layers in biomagnetites are shaped by a slow extraction which can be directed via electric and electromagnetic cellular processes. When the time of bird migration comes, migration intention can induce a genetic program in brain cells of migratory birds, which initiates and operates the development of biomagnetites. During migration, layers of biomagnetites and phases of non-conductive electrons in the layers are shaped by the current magnetic vector potential field of Earth (see Fig.3a). These polarized states of non-conductive electrons – which are fixed in current layers in biomagnetites – can play the main role in magnetic information storage.

During the return of a bird, previously fixed vector potentials of Earth's magnetic fields can induce an Aharonov-Bohm type of oscillation in the fixed layer of biomagnetites (see Fig.3b). Therefore, the electric resistance of biomagnetites oscillates. Localized – previously fixed and spin-modulated – non-conductive electrons act as scattering sites for the mobile electrons. Namely, oscillations of dephasing non-conductive (fixed) electrons can have an influence on conductive mobile electrons in biomagnetites. Then, there is a coherent transport of mobile electrons' spins into surrounding semiconductor protein molecules. Both the effects of electric resistance oscillations in biomagnetites and the transport of spins into proteins can induce conformational changes in organic proteins at the free rotations. Eventually, conformational changes in organic molecules are amplified within cells, and then among cells, which can direct the movement of migratory birds. In this concept, biomagnetites work as ″devices″ of spin-modulated information storage.

Biomagnetites can perceive the Earth's magnetostatic or vector potential fields while the layers of them have taken shape, although the magnetostatic interaction between the magnetic nanoparticles is negligible in the 2D nanoscale limit [58]. Accordingly, biomagnetites' magnetic state can be manipulated separately from the state of neighboring biomagnetites. Consequently, the vector potential of the Earth's magnetic field (which cannot be shielded) plays the main role, and there is no need for a strong external magnetic field to have a mechanical effect (closing or opening ion gates) on biomagnetites.

Wernsdorfer and Sessoli have observed an Aharonov-Bohm type of oscillation in magnetic molecular clusters, analogous to the oscillations as a function of the external flux in a SQUID ring [63]. Their opinion is: "A great deal of information is contained in these oscillations both about the form of the molecular spin Hamiltonian and about the dephasing effect of the environment". This spin "memory" concept about biomagnetites can guarantee a great amount of information and an operating system, which is needed for the migratory bird to find its way home from a distance of some thousand kilometers.

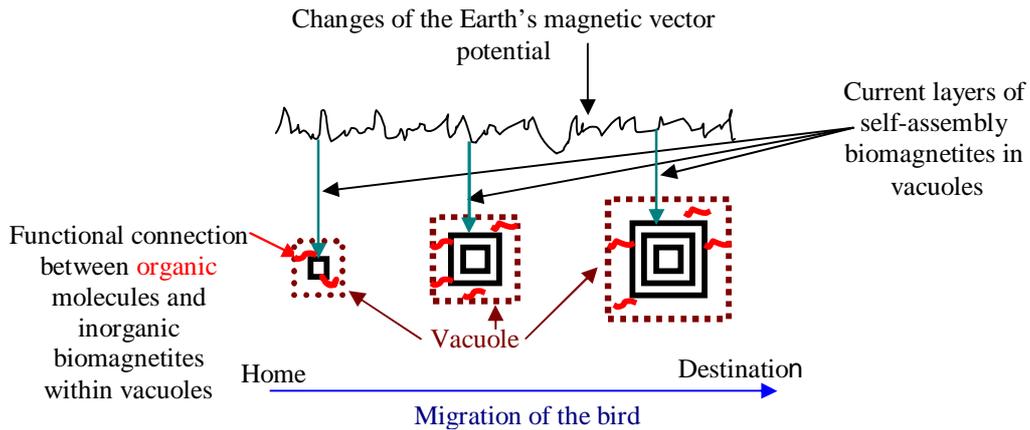

**Fig. 3a.** When the time of bird migration comes, migration intention can induce a genetic program in brain cells of migratory birds, which initiates and operates the development of biomagnetites in vacuoles. During migration, layers of biomagnetites and phases of non-conductive electrons in the layers are shaped by current magnetic vector potential field of Earth (Fig.1a). These polarized states of non-conductive electrons – which ones are fixed in current layers in biomagnetites – can play the main role in magnetic information storage.

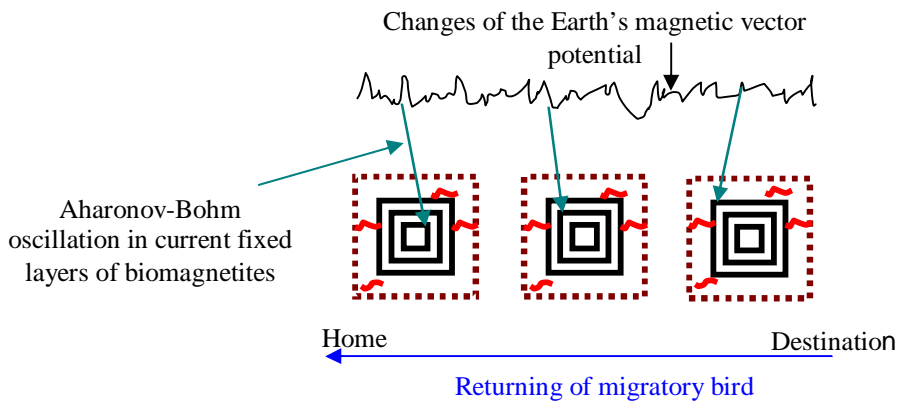

**Fig. 3b.** During return of bird, previously fixed vector potentials of Earth's magnetic fields can induce an Aharonov-Bohm type of oscillation in the fixed layer of biomagnetites. Therefore, the electric resistance of biomagnetites oscillates. Localized – previously fixed and spin-modulated – non-conductive electrons act as scattering sites for the mobile electrons. Namely, oscillations of dephasing non-conductive (fixed) electrons can have an influence on conductive mobile electrons in biomagnetites. Then, there is a coherent transport of mobile electrons' spins into surrounding semiconductor protein molecules. Finally, the oscillation of the electric resistance in biomagnetites as well as the transport of spins into proteins can induce conformational changes in organic proteins at the free rotations.

## 7. Signal amplification process of spin information

In reality, many biomagnetites and cells take part in these cooperative processes, but static magnetic forces of biomagnetites and environment have no important role in it while magnetic vector potentials do. This concept does not need the existence of biomagnetites in most cells of the bird's brain. Fixed vector potential information in biomagnetites is converted into different electrical vibration signals in cells.

Aharonov-Bohm oscillations of electric resistance in biomagnetites and also transport of spins into semiconducting proteins can change conformations of organic molecules, which are in direct connection with biomagnetites. Then, conformational changes can oscillate and these oscillations which appear as conformons - in a polar biological system - generate cellular electromagnetic fields around themselves, which can mediate long-range interactions and also signal amplifying processes [20, 21].

The living matrix is a structural and energetic continuum. Every cell contains a cytoskeleton that is connected, across the cell surface, with the extracellular connective tissue matrix. Structural components of this system include the connective tissue, cytoskeletal, musculoskeletal, and genetic networks. Thus, propagation of conformational changes can expand by this network in cells.

Coupled oscillations, resonant transfer, and electrodynamic coupling allow energy and information to flow through the network. Therefore, electromechanical, electrochemical, or electromagnetic signals can regulate the signal-amplifying processes within and between cells.

## 8. Biomagnetites in the human brain

The Aharonov-Bohm effect can play important roles in biomolecular communication. This effect is able to connect the different weak electric, magnetic, and electromagnetic signal processes in various cells. Magnetic informational processes can serve unconscious navigation of living creatures. It seems that humans also have an innate unconscious sense of direction by direction-sensitive cells [57]. However, biomagnetites must also have functional roles in human place cells, for if not, why are there billions of biomagnetites (*among others in hippocampus where direction-sensitive place cells also exist, and vision is not necessary for normal firing of hippocampal place cells*) with perfect structure in the human brain cells?! [17, 18, 55].

## 9. Summary

Living cells can generate and use electric, magnetic, electromagnetic, and acoustic waves (conformons), and convert from one form into another. The biological regulation of biomagnetite formation can be achieved by surrounding organic protein molecules and coherent cellular electromagnetic fields (biophotons).

In this paper we have hypothesized that spin-modulated resistance can work in biomagnetites through the Aharonov-Bohm effect. Layers in biomagnetites are shaped by a slow extraction that is regulated by electric and electromagnetic cellular processes. When the time of a bird's migration comes, it can induce a genetic program in brain cells of the migratory birds, which initiates and operates the development of biomagnetites. During migration, biomagnetic crystal layers and phases of non-conductive electrons within layers are shaped by current magnetic vector potential field of the Earth. During the return of the migratory birds, earlier fixed vector potentials of Earth's magnetic fields can induce an Aharonov-Bohm type of oscillation adequately and then fix biomagnetites layer. Therefore, the electric resistance of biomagnetites will exhibit oscillations. Localized – previously fixed and spin-modulated – non-conductive electrons act as scattering sites for the mobile electrons. Namely, oscillations of dephasing non-conductive (fixed) electrons can have an influence on conductive mobile electrons in biomagnetites. Then, there is a coherent transport of mobile electrons' spins into surrounding semiconductor protein molecules. Both effects, electric resistance oscillations in biomagnetites and transport of spins states into proteins, can induce conformational changes in the surrounding organic proteins in the free rotations. Eventually, conformation changes are amplified by signal processes in cell, and then among cells, which can then direct the movement of a migratory bird. Within this concept, biomagnetites work as ″devices″ performing electron spin-modulated information storage. Here we also emphasize that biomagnetites must have functional roles in the human place cells.